\UseRawInputEncoding
\documentclass[a4paper,10pt,oneside]{article}
\usepackage[utf8]{inputenc}
\usepackage{
    icad2024,
    amsmath,
    epsfig,
    times,
    url,
    hyperref,
    accessibility,
    csquotes
    }
\usepackage[T1]{fontenc}
\usepackage[dvipsnames]{xcolor}
\usepackage[
    backend=biber,
    style=numeric-comp,
    maxcitenames=5,
    maxbibnames=1,
    sorting=none,
    backref=false
]{biblatex}
\hypersetup{
    colorlinks=true,
    linkcolor=RoyalBlue,
    filecolor=RoyalBlue,      
    urlcolor=RoyalBlue,
    citecolor=RoyalBlue,
}

\title{Spin-Wave Voices: Sonification of Nanoscale Spin Waves\\ as an Engagement and Research Tool}

\twoauthors{S. Pile, O. Lesota, S. D. Peter, C. Humer} {Johannes Kepler University Linz \\ Altenberger Strasse 69\\ 4040 Linz, Austria  \\ {\tt santa.pile@jku.at}\\ {\tt oleg.lesota@jku.at}\\ {\tt silvan.peter@jku.at}\\ {\tt christina.humer@jku.at}}
{M. Gasser} {University of Applied Arts Vienna \\ Institute of Arts and Society \\ Georg-Coch-Platz 2 \\ A-1010 Vienna, Austria  \\ {\tt martin.gasser@uni-ak.ac.at} \\ \\Muse Group \\ Limassol, Cyprus \\ {\tt martin.gasser@mu.se}}

\begin{document}
\ninept
\maketitle
\begin{sloppy}
\begin{abstract}
Magnonics is an emerging research field that addresses the use of spin waves (magnons), purely magnetic waves, for information transport and processing. Spin waves are a potential replacement for electric current in novel computational devices that would make them more compact and energy efficient. The field is yet little known, even among physicists. Additionally, with the development of new measuring techniques and computational physics, the obtained magnetic data becomes more complex, in some cases including 3D vector fields and time-resolution. This work presents an approach to the audio-visual representation of the spin waves and discusses its use as a tool for science communication exhibits and possible data analysis tool. The work also details an instance of such an exhibit presented at the annual international digital art exhibition Ars Electronica Festival in 2022.
\end{abstract}

\section{Introduction}
\label{sec:intro}
\textit{Spin-Wave Voices} is an art installation, which was presented at the digital art exhibition Ars Electronica Festival \enquote{Welcome to Planet B}\footnote{\href{https://ars.electronica.art/planetb/en/spin-wave-voices/}{https://ars.electronica.art/planetb/en/spin-wave-voices/}}
. This exhibition allowed the wide public to explore the innovative technology of future computational devices: the spin waves. The processes behind the capabilities of modern digital devices happen on micro- and nanoscopic scales and at incredible speeds, making them impossible to observe with the naked eye and difficult to comprehend~\cite{Gelsinger2001, Nikonov2022}. Spin waves (magnons) are considered to be a potential replacement for electrons for data processing in novel computational devices~\cite{Chumak2022,Wang2018}. The \textit{Spin-Wave Voices} exhibit allowed visitors to experience the future technology through visualization and auditory display (AD) in a playful, easy-going way. Furthermore, the people from the field were allowed to get insights into the data and spin-wave behavior.

Schematic of the installation is shown in Fig.~\ref{fig:cube1}. Every visitor had a chance to excite spin waves in 5 differently shaped microstrips (a) by pressing corresponding pedals (b), as described in more detail in Section ~\ref{sec:Installation}. The waves were visualized and sonified in a way, that the spatial differences of the waves caused by the variation of the shapes of the microstrips were easily recognizable by visitors.

\begin{figure}
    \centering
	\includegraphics[width=0.5\textwidth]{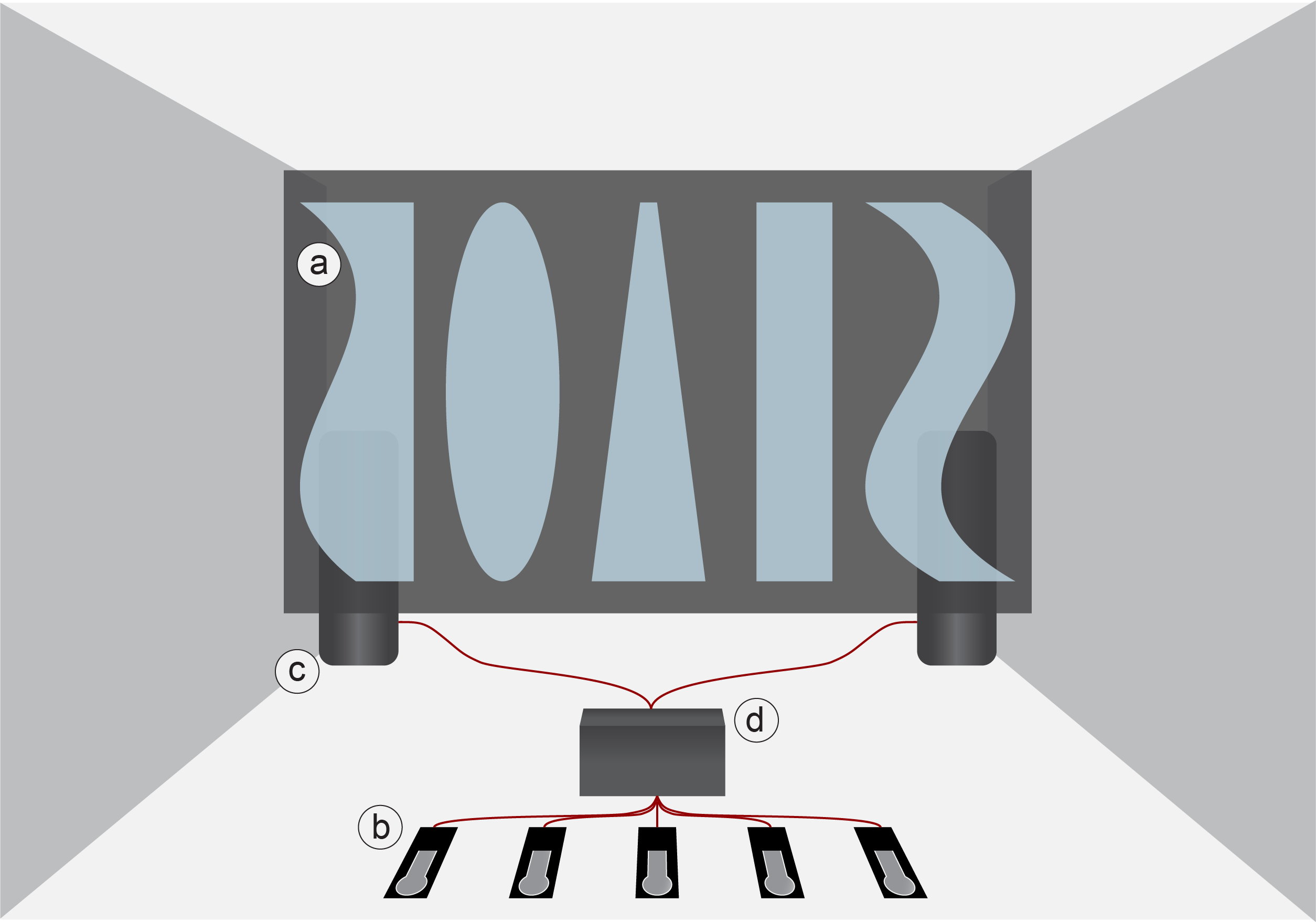}
	\caption{\textit{Spin-Wave Voices'} Space B (see Section~\ref{sec:Installation} and Fig.~\ref{fig:cubePlan}) schematic representation: (a) screen, (b) pedals, (c) loudspeakers, and (d) projector.}
	\label{fig:cube1}
\end{figure}

\subsection{Spin Waves for Future Computing}
\label{sec:introSWs}
To prevent the energy crisis of the near future, one needs to search for new approaches in technology development already now. One of the biggest contributions to that nowadays is computing. Modern computational technologies, such as artificial intelligence (AI), have provided enhanced features for modern devices and data processing, but as well the total energy consumption by general-purpose computing continues to grow exponentially and is doubling approximately every 3 years while the world’s energy production is growing only linearly, by approximately 2\% a year (see Fig. 8 in~\cite{Interim2021}).

One can approach this problem, e.g. optimizing computing strategies, as is already actively researched for AI, so-called AI research strategies~\cite{Verdecchia2023}. Another way would be making computing devices more energy efficient. Magnonics is a promising research field in this regard, as it employs spin waves (magnons), purely magnetic waves~\cite{Rana2019}, for information transport and processing. Among other advantageous features, magnonic devices can be scaled down to atomic dimensions, can operate in a wide frequency range up to hundreds of THz, are able to process data at temperatures spanning from ultra-low to room temperature. Additionally, spin waves can transfer data without Joule heating~\cite{Chumak2014} and can make unconventional computing possible due to their nonlinear properties~\cite{Chumak2022}. Moreover, there has been increasing interest in studying spin waves, in various geometries, including curved structures, as 3D curvature can introduce unique magnetic properties and novel phenomena due to the interplay of curvature, topology, and magnetic interactions~\cite{Streubel2016,Fischer2020}.

The research of magnonics, especially when it comes to the investigation of patterns in spin wave behavior, can often be impeded by an extensive complexity of the data, either measured or obtained from micromagnetic simulations~\cite{Donnelly2017}. Analysis of complex data requires the use of sophisticated visualization techniques in order to reduce the cognitive load on the analyst while keeping the visualization representative of the phenomenon~\cite{Crameri2020}. 

\begin{figure}
    \centering
	\includegraphics[width=0.32\textwidth]{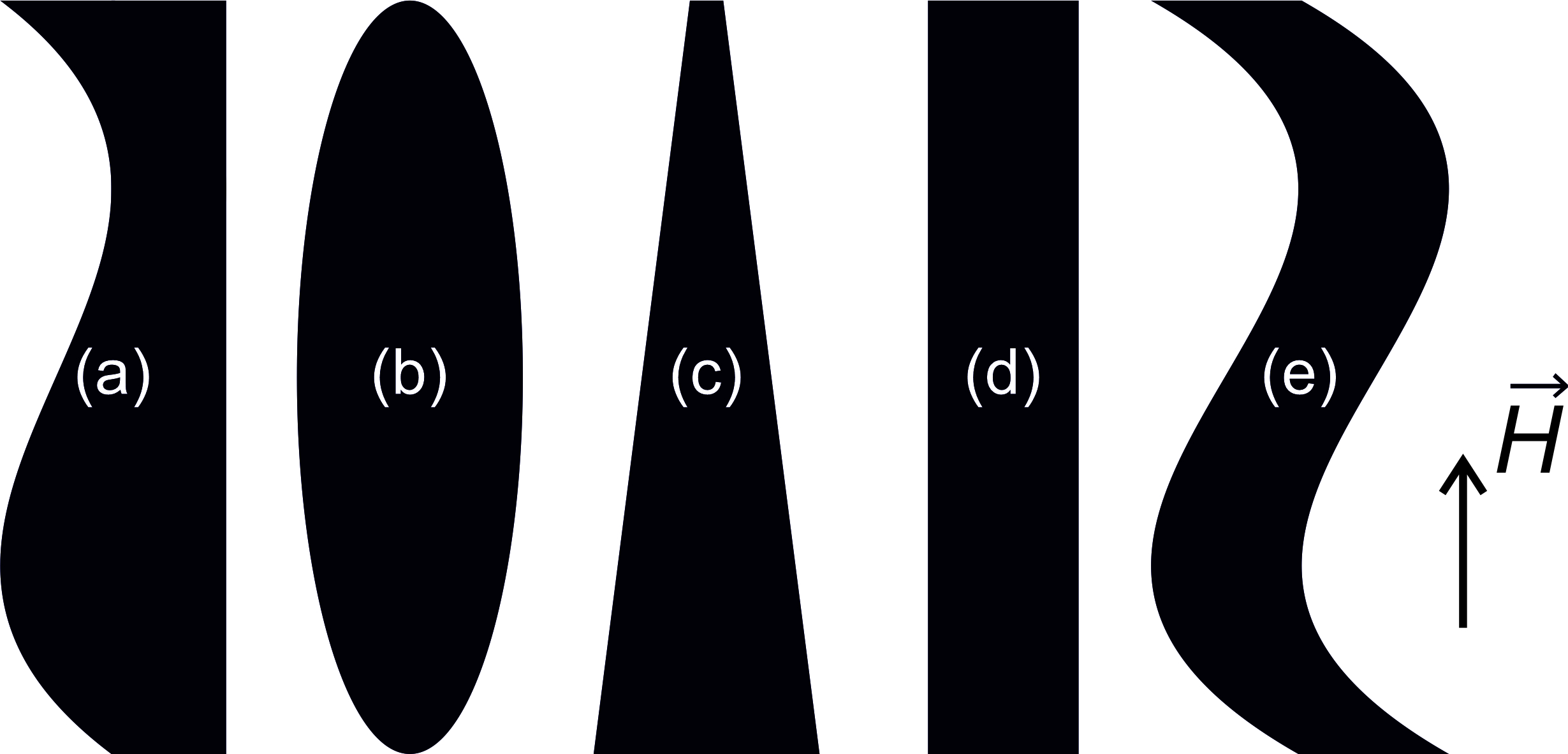}
	\caption{Strip shapes used for auditory display (AD): (a) vase, (b)~ellipse, (c) pyramid, (d) strip, and (e) wave.}
	\label{fig:shapes}
\end{figure}

\subsection{Complex Data Visualization}
\label{sec:introVis}
Measurement or simulation data of physics phenomena, like spin waves, are too complex to understand in their raw format. Scientific visualization techniques are frequently applied to facilitate the understanding and interpretation of spatial (scalar) fields~\cite{schroeder_visualization_2006,telea_data_2014}. Such datasets include fixed spatial positions for the scalar values, leaving only a small subset of effective channels for encoding the scalar value itself~\cite[p. 102]{munzner2015visualization}. Among the more effective channels for encoding a scalar value under these restrictions are depth (i.e., 3D position) -- if the scalar field is a 2D plane -- and color. The use of depth for encoding scalar fields results in a 3D visualization that comes with drawbacks like occlusion, distortion, or multiple viewpoints~\cite[pp. 120-122]{munzner2015visualization}. Therefore, the use of 3D visualizations must be justified and the benefits must outweigh the drawbacks. One such benefit is to support people in building a mental model of 3D phenomena~\cite[p. 124]{munzner2015visualization}. In particular, we can use waves on a water surface as a metaphor for spin waves to aid people in building a mental model about them. We can use alternative 3D representations (e.g., using slices of the data represented as a ridgelines plot~\cite{wilke_fundamentals_nodate}) to avoid occlusion.

Time-dependent data represents another challenging case in visualization. Plotting temporal data over the time-axis (i.e., animated transition between timesteps) is a natural way of representing such data. However, animated data requires viewers to keep previous states in memory, increasing people's cognitive load~\cite[p. 132]{munzner2015visualization}. The risk of missing certain patterns due to selective attention might result in the necessity to re-watch an animation multiple times. This hampers the ability to analyze such complex data visually. Ways to augment the analyst and to help them cope with complex data include splitting complex visualization into multiple views and utilizing their human senses beyond vision. In particular, the human auditory system is designed to \enquote{recognize temporal changes and patterns}~\cite{enge_open_2024}.

Although analyzing animated data is less accurate, animations are also shown to be more enjoyable and exciting to viewers~\cite{robertson_effectiveness_2008}. Visualizations must therefore be carefully selected under consideration of the target user group and the task goals in mind. Domain experts (e.g., physicists), for example, need to analyze data to discover new insights~\cite[p. 47]{munzner2015visualization}. For non-experts (e.g., exhibition visitors), on the other hand, visualizations can be used to present a topic and educate people or increase their awareness of a topic~\cite[pp. 47-48]{munzner2015visualization}. For the latter, a visualization should be designed to be informative and engaging, but not too complex. 

\subsection{Auditory Display and Sonification}
\label{sec:introAD}
The ability of AD to provide a better understanding, or an appreciation, of changes and structures in the sonified data has been attracting rapidly growing attention over the last years. One of the fields that benefited from the sonification and auditory display of complex data is astronomy~\cite{Zanella2022}. The reason for that is auditory display being one of the most appropriate approaches when the data has complex patterns and changes in time~\cite[chapter~2]{Hermann2011}.

With the development of measuring techniques~\cite{Donnelly2017,Weigand2022} and computational physics, e.g. micromagnetics~\cite{Abert2019,Leliaert2019}, the field of magnonics meets similar challenges as in astronomy. The obtained data, which needs to be analyzed, becomes more complex, in some cases including 3D vector fields and time-resolution. Adding AD to visualization tools could benefit from human auditory perception, which allows for monitoring and processing multilayered auditory data. Additionally, the ability of humans to learn from sonified data and being able to notice that \enquote{something is wrong}~\cite[chapter~1]{Hermann2011} can open new opportunities for getting unexpected insights from the data.

Another important aspect of the AD included in the data representation is education and accessibility of the new research to a broader audience. Magnonics is a relatively new research field, even though it might be one of the possible solutions for the energy crisis. Including AD in an exhibition, which is aimed at raising the awareness of the public about the problem and its possible solutions, as will be shown further, significantly increased audience interest.

\subsection{\textit{Spin-Wave Voices} Objectives}
\label{sec:objectives}
In this work, an approach to combining data sonification with interactive visualization of spin waves, excited in confined Ni$_{80}$Fe$_{20}$ (Py) microstrips of different shapes, is presented. In particular, the influence of the shape of a structure on spin-wave patterns during a transient response (from the initial moment up to the stable state) of the system on uniform microwave (MW) field excitation is illustrated. Even though excitation conditions for all shapes are exactly the same, one can experience different spin-wave patterns depending on the shape. This can show one of many unique properties of spin waves, which make them a promising data carrier for future computational devices and trigger fruitful discussions leading to better awareness of the phenomena. A prototype developed following our approach was named \textit{Spin-Wave Voices} and has been evaluated in the science-communication framework at a digital art exhibition Ars Electronica Festival 2022 \enquote{Welcome to Planet B}\footnote{\href{https://ars.electronica.art/planetb/en/spin-wave-voices/}{https://ars.electronica.art/planetb/en/spin-wave-voices/}}. The \textit{Spin-Wave Voices} installation was aiming to:
\begin{enumerate}
    \item raise public awareness of the need for energy-efficient computational technologies, e.g. spin waves;\label{obj:1}
    \item give visitors an opportunity to imagine nano-scale ultrafast processes akin to ones behind modern and future computational technologies;\label{obj:2}
    \item provide an accessible immersive audio-visual experience illustrating the basic concept of spin waves in computing;\label{obj:3}
    \item explore the potential of sonification as a tool for analysis of spin-wave data.\label{obj:4}
\end{enumerate}

In the following sections detailed description of technical details of the project, its presentation and reception at the festival are discussed. Additionally, \textit{Spin-Wave Voices} was presented at two research groups, who specialize on magnetic phenomena, and at the IEEE International Magnetics Conference 2024 (INTERMAG 2024) as a contributed talk, which is also discussed in more detail in Section~\ref{sec:demoSpecField}.

\section{Technical Details of the Project}
\textit{Spin-Wave Voices} included several components, such as simulation of the real physical system in order to generate data, interactive visualization and sonification of the data, and synchronization of the latter to represent a transient response of spin waves on the excitation, which is controlled by a visitor. In the following subsections, each of the components is described in more detail.

\subsection{Spin-Wave Data Source}
\label{sec:SWVDataSourse}
As a data source results of micromagnetic simulations of spin-wave transient response (i.e. dynamics from the beginning of the excitation till the steady state is achieved) in Ni$_{80}$Fe$_{20}$ microstrips of various shapes are used. Spin waves in the structures are excited by applying an in-plane static magnetic field $\mu_0H=87$~mT in a direction indicated in Fig.~\ref{fig:shapes}, and an out-of-plane periodic magnetic field at excitation frequency $f_\mathrm{MW}=9.4$~GHz.

Micromagnetic simulations were performed using MuMax3~\cite{mumax} with the following parameters~\cite{Pile2022}: frame size that included sample material and an empty space of $1.5\times5~\mu m$, a thickness of 25~nm, saturation magnetization $M_\mathrm{s}=730$~kA/m, Gilbert damping parameter $\alpha=0.008$, exchange stiffness constant $A_\mathrm{ex}=13\times10^{-12}$~J/m, and g-factor=2.12. As a result, 20 frames per excitation period of the spatially resolved magnetization configuration over the first 50 excitation periods were recorded from the simulations. As a result of the simulations for each shape, 1000 magnetization vector field configurations were obtained, depicting an evolution in time of the magnetization dynamics. Each configuration's spatial distribution of the out-of-plane magnetization component was extracted as a 2D array of data (spin-wave pattern). This resulted in 1000 data arrays $500\times 150$ for each shape, which were further used for the visualization and AD.

The shapes were chosen in a way that with the same excitation parameters, such as external static and excitation periodic magnetic field, the spin-wave patterns would differ significantly. The resulting shapes are shown in Fig.~\ref{fig:shapes}. With the chosen parameters a uniform excitation is observed in the ellipse shape (b), and spin waves are excited in the other four shapes with different wave lengths.

\begin{figure}
    \centering
	\includegraphics[width=0.45\textwidth]{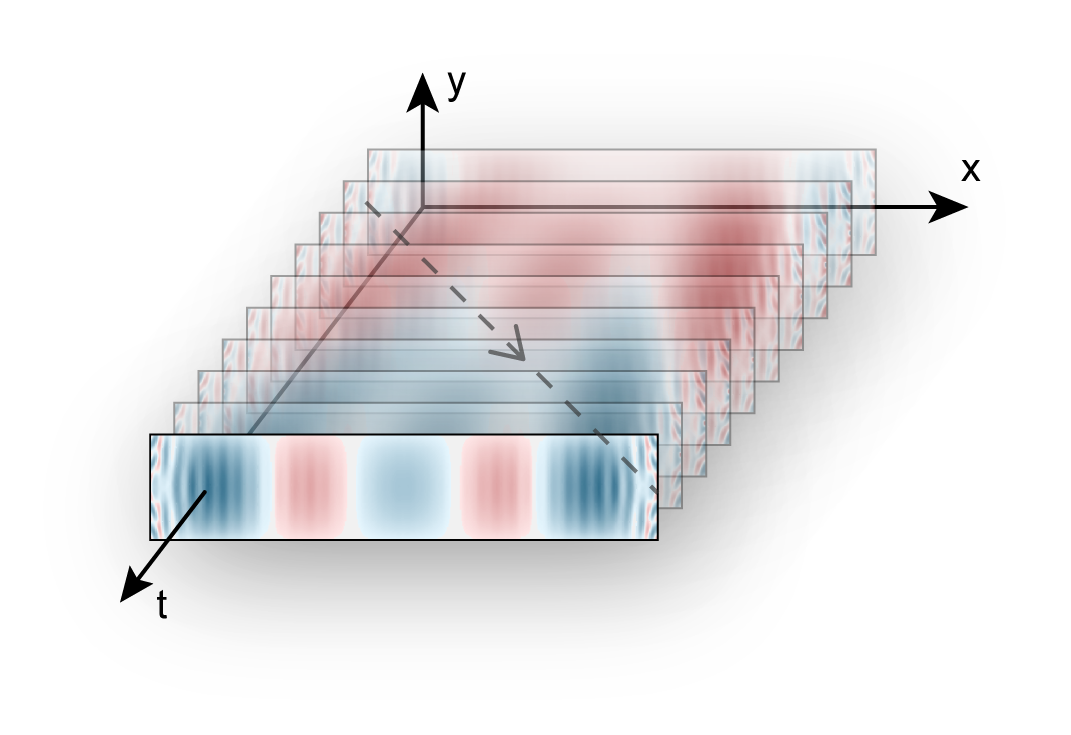}
	\caption{A stack of the simulated images over one excitation period for a rectangular strip shape shown in  Fig.~\ref{fig:shapes}~(d).}
	\label{fig:imageStack}
\end{figure} 

\subsection{Interactive Visualization}
\label{sec:interactiveViz}
The interactive visualization for \textit{Spin-Wave Voices} was created using Cables.gl\footnote{\href{https://cables.gl/p/WPjgr6}{https://cables.gl/p/WPjgr6} - for the Editor view only.} (cables). Cables is a free online tool for creating interactive visual web content. It was developed by the Berlin-based studio undev\footnote{\href{https://undev.studio}{https://undev.studio}} initially as a development tool for interactive WebGL and Web Audio projects. It uses a node-based interface, allowing for quick prototyping.

For visualization, the generated data arrays (as described in Section~\ref{sec:SWVDataSourse}) were plotted as heatmaps, in which color hue indicates a sign of the out-of-plane magnetization [blue: negative, red: positive in Fig.~\ref{fig:imageStack}, and brown for positive in Fig.~\ref{fig:visAllShapes}~(a)], and color brightness indicates its magnitude. The heatmaps were saved as images, imported in cables as a texture and converted into a 3D surface mesh, where the vertices are displaced with the pixel's brightness values from the texture. Then the mesh is converted into an array of horizontal lines, as can be seen in Fig.~\ref{fig:visAllShapes}~(b).

The goal of the visualization is to showcase the transient response of a micro-structure to uniform MW field excitation. This is achieved by leveraging a three-dimensional visualization, described above, combined with an animation. The generated visualization represents the transient response of a micro-structure by plotting the out-of-plane magnetization component on the outline of the structure at every moment in time during excitation. The plots are then used to create an animation. The out-of-plane component is encoded both with a diverging color scheme, which indicates the direction of the out-of-plane component, and as elevation/depression out of the plain of the structure outline, indicating the magnitude of the magnetization.

As a result, the visualization appears as a surface: flat before the excitation (out-of-plain magnetization is zero) and, after excitation is applied colored, dynamically undulating with elevations and depressions demonstrating the transient behavior. The animation is controlled by the viewer: at any point, they are able to restart the visualization/excitation of the spin waves and observe the evolution of the process anew. To be perceivable by a human the process is visualized at a rate of half of the excitation period per second.

\begin{figure}
    \centering
	\includegraphics[width=0.45\textwidth]{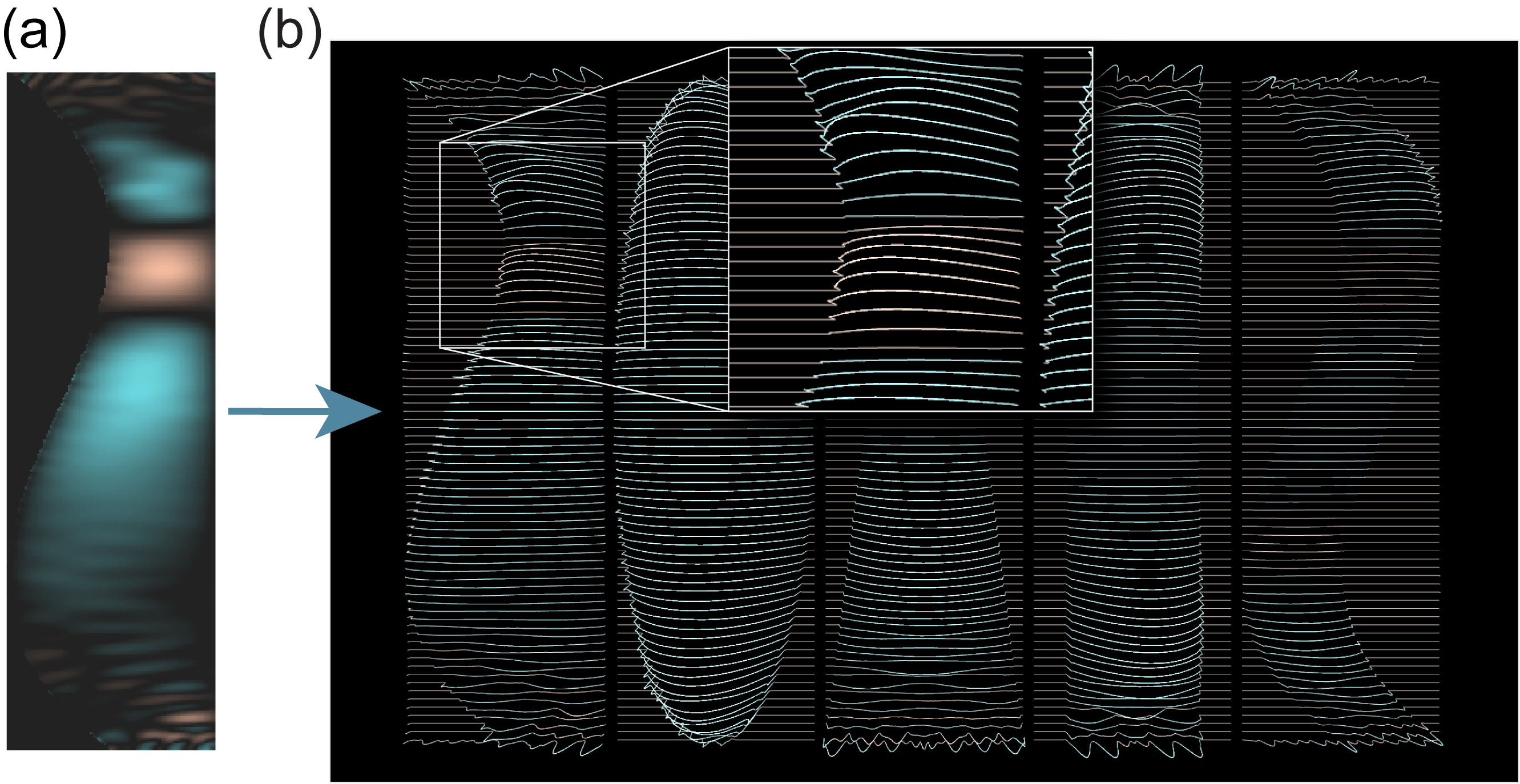}
	\caption{(a) An example of a heatmap image, used for the visualization. (b) Screenshot of the visualization with an enlarged section to show the ridge lines in more detail.}
	\label{fig:visAllShapes}
\end{figure}

\subsection{Sonification of Spin-Wave Data}
\label{sec:sonAD}
The simulation data is two-dimensional, time-dependent wave data and thus suitable for audification-type sonification~\cite[chapter~12]{Hermann2011}, with considerations given to the spatial and frequency characteristics of these measurements. 

In order to convert the two-dimensional data generated by micromagnetic simulations to one-dimensional waveform data, Scanned Synthesis~\cite{Verplank}, a descendant of Wavetable Synthesis, is adapted for \textit{Spin-Wave Voices}. Wavetable Synthesis generates audio wave forms by periodically sampling another wave form at a given frequency. This way, the timbre and the fundamental frequency of the resulting wave form can be controlled independently. Scanned Synthesis extends Wavetable Synthesis in two respects. (1) the wave table can be a low-frequency dynamic system and (2) the wave table can be two-dimensional. In order to sample a one-dimensional audio signal from a two-dimensional dynamic system, a periodic sampling path is defined, and the data is upsampled to match the audio sample rate.

Audio signals are sampled at a predefined audio sample rate, e.g. $44.1$~kHz, from $500\times150$~pixel image streams at a predefined image frame rate, e.g. $10$~fps. Periodic audio-rate sampling path $(x, y, t)$ through the spatio-temporal data grid illustrated in  Fig.~\ref{fig:imageStack} with the gray dashed line with an arrow indicating the sampling direction is predefined. The sampling period of the path defines the fundamental frequency of the resulting musical audio signal while the data itself (which can be also observed in the corresponding visualization) creates a dynamic, pulsating texture that gives each shape a certain timbre. To generate the audio data, the data is linearly interpolated at the locations defined by the sampling path. Because only one audio sample value per resampled image is calculated and to avoid wasting resources, only the path is resampled and not full images for each audio sample. The latter is achieved by linear interpolation from the image data \enquote{on the fly}, by calculating one audio sample value for each $(x, y, t)$ value.

The simulation data is sonified from five different shapes, plus the installation setup resembles a piano-like instrument, therefore, the frequencies from a pentatonic scale (98 Hz, 110 Hz, 123 Hz, 165 Hz, 185 Hz) were chosen as fundamental frequencies that define the period of the sample paths. The resulting audio data is written to files and imported into an Ableton Live's Looper instrument. As the simulation data consists of a transient and a steady-state segment, the times of the segment boundaries are identified and loop points are set accordingly. When a pedal is pressed, the transient segment is played back once and the steady-state portion is repeated as long as the corresponding pedal is pressed. As the floor pedals are set up from left to right, the corresponding loops are panned from left to right to reflect shapes' spacial arrangement.

\subsection{Hardware + Software}
As schematically shown in Fig.~\ref{fig:cube1} the physical realization of the installation consists of a screen (a), a projector (d), a sound system (c), five foot pedals (b) that can be used by the visitors to trigger the visualization and the sonification (see additionally Fig.~\ref{fig:cubePlan}), and a regular PC laptop for running the visualization and sonification processes (not shown in the figures). As an input devices a standard sostenuto pedals connected to an Arduino microcontroller that translates the voltage changes to MIDI messages were used. The pedals were placed in a line on the floor, spaced widely enough to enable pressing no more than two pedals by a single visitor. These messages are then sent to the PC running Ableton Live for the sonification part, and a web browser running a cables.gl patch for the visualization\footnote{\href{https://github.com/Spin-Wave-Voices/SWV-App}{https://github.com/Spin-Wave-Voices/SWV-App} - offline simplified version (no pedals).}.

The sounds were pre-rendered using a Python script that reads the simulation data, implements the synthesis algorithm described in Section \ref{sec:sonAD}, and generates the audio files. Whenever a visitor pressed and held one of the foot pedals, the corresponding sound was played back by Ableton Live. Each sound sample starts with the sonification of the transient initial phase, where the spin-wave stable state is formed under the uniform excitation. When steady state is reached a small sample segment is looped. The browser-based visualization patch receives the same MIDI message and synchronously renders the data in real-time, starting from the same relaxed state towards a (looped) steady state. The visualization was projected on a screen in front of the visitors. See Fig.~\ref{fig:flowChart} for an overview of the data flow. Releasing the pedal at any moment sends another MIDI message and stops the playback and rendering and also resets to the initial state. This setup encourages close and sustained attention as visitors need to continuously hold the pedal to keep the sonification and visualization changing.

\begin{figure}
    \centering
	\includegraphics[width=0.28\textwidth]{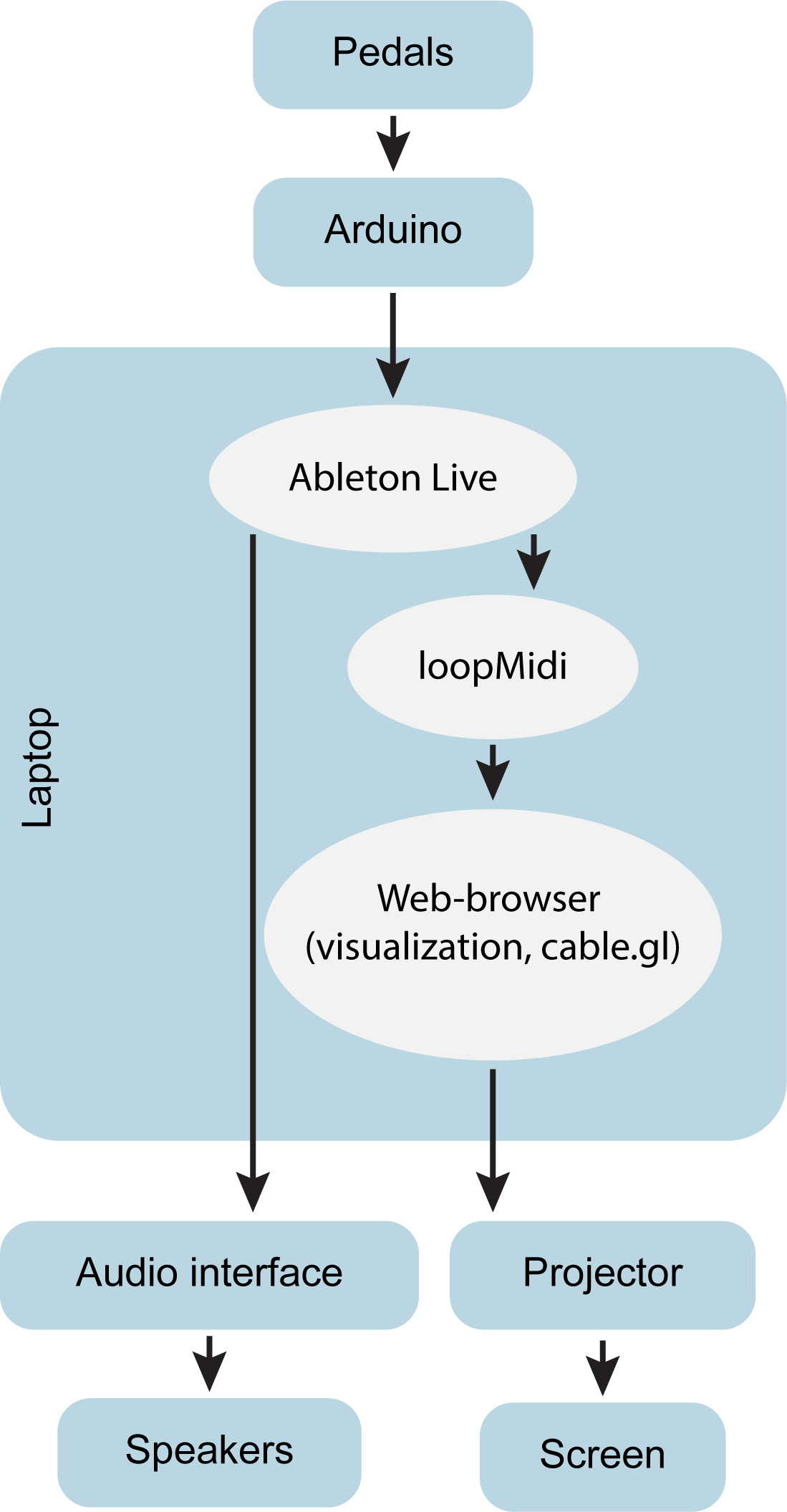}
	\caption{Technical workflow of the audio-visual installation.}
	\label{fig:flowChart}
\end{figure}

\section{\textit{Spin-Wave Voices} Experience}
\label{sec:Installation}
The design of \textit{Spin-Wave Voices} is driven by both science communication and artistic goals (see Section~\ref{sec:objectives} for more detail) and strives to let the two aspects synergize and support each other. The science communication goals include raising public awareness of computational and energy-efficiency limitations of modern computing technologies, as well as illustrating the concept of spin waves, a phenomenon that could form the basis for new more efficient and effective computing. The artistic goal of the exhibit is to emotionally and sensually convey the complexity and importance of the nano-scale ultra-fast processes powering contemporary computational technologies. The exhibit is split into two spaces, A and B (Fig.~\ref{fig:cubePlan}), each aimed at a specific set of goals.

\subsection{Space A: Science Communication}
\label{sec:spaceA}
Space~A, which is encountered by a visitor upon entering the exhibit, serves the science communication goals, providing context to the audio-visual experience in Space~B. At all times a guide is present in the room offering an introduction to new visitors. The brief introduction covers the following points:
\begin{itemize}
    \item Modern computing is current-based and is about to hit its limits in terms of energy efficiency (due to the heat produced by electric current going through conductors) and size;
    \item Spin waves is a phenomenon that could replace current in computing allowing smart devices to become faster, more compact and energy efficient, including a short explanation of what spin waves are;
    \item The possibility to change spin-wave patterns by just changing the shape of a wire, which is shown at the exhibit, demonstrates the flexibility of spin waves and, among other already mentioned advantages, their potential for the realization of non-conventional computing;
    \item In the next room (Space B) the visitors can interact with the audio-visual installation by means of foot pedals. Pressing a pedal starts the simulation of a periodic magnetic field exciting spin waves in a microstrip of the corresponding shape.
\end{itemize}
To help make the introduction more engaging the room features a microscope allowing to demonstrate the scale of the microstrips by showing a rectangular strip [Fig.~\ref{fig:shapes}~(d)] next to a human hair (see Microscope image in Fig.~\ref{fig:cubePlan}). The room also has a scientific poster placed on the wall allowing more motivated visitors to delve deeper into the physics behind the spin waves. In addition, the walls are decorated with two cleanroom suits - a special kind of clothing worn in a cleanroom (space, which maintains a very low amount of airborne particulates) while producing microstrips. A cleanroom environment is required to prevent, e.g. dust particles, which are usually larger than the strips, from interfering with the production process. After the introduction, visitors are invited to experience the audio-visual display in Space~B.

\subsection{Space B: Diving into Spin Waves}
\label{sec:spaceB}
The artistic goal of the exhibit is approached by making spin waves audibly and visually perceivable for a human, which is impossible under normal circumstances due to the scale and speed of the phenomenon. Therefore, the exhibit features Space~B, a special space allowing visitors to overcome limitations of human perception and experience what spin waves could look and sound like, see Space~B in Fig.~\ref{fig:cubePlan}. The room has no lights except for the minimum required for the visitors' safety. One of the walls in the space is occupied by a large screen [Fig.~\ref{fig:cube1}~(a)] where visual representations of five different microstrip structures are projected (see Section~\ref{sec:interactiveViz}). Loudspeakers [Fig.~\ref{fig:cube1}~(c)] are placed behind the screen to reinforce the sensation of the sound coming from the strips. About 3 meters in front of the screen five foot pedals, one for each projected microstrip, are set up [Fig.~\ref{fig:cube1}~(b)]. By pressing a pedal visitors are able to start the simulation of spin-wave excitation in the respective strip. As a result, the strip visualization starts pulsating following the simulated patterns of magnetization (see Sections~\ref{sec:SWVDataSourse} and \ref{sec:interactiveViz}), while the audio representation of the process is played in parallel (see Section~\ref{sec:sonAD}). A video recording of the exhibit showing the path through spaces A (without a brief introduction by a guide) and B including an example of interaction with the pedals is available\footnote{\href{https://youtu.be/DVGgFnQUK1g}{https://youtu.be/DVGgFnQUK1g}}.

The audio-visual sequence of each strip is unique: gradually evolving from the initial state (no spin waves excited) to a distinct stable state (the spin-wave pattern is stable). The screen recordings of the sequences corresponding to each strip are available online: vase\footnote{\href{https://youtu.be/a8Ruy1NVoG0}{https://youtu.be/a8Ruy1NVoG0}}, ellipse\footnote{\href{https://youtu.be/bt7djw2yBnA}{https://youtu.be/bt7djw2yBnA}}, pyramid\footnote{\href{https://youtu.be/CtKyRkgrZQE}{https://youtu.be/CtKyRkgrZQE}}, rectangular strip\footnote{\href{https://youtu.be/tpxx9haPGOE}{https://youtu.be/tpxx9haPGOE}}, wave\footnote{\href{https://youtu.be/ucMj71vy8bg}{https://youtu.be/ucMj71vy8bg}}. The audio-visual display with its loud sound and pulsating imagery turns a humble micro-scale element, one whose countless barely noticeable counterparts serve as building blocks for modern smart devices, into a force to be reckoned with. The contrast between seeing a microstrip being hundred times smaller than the width of a human hair and experiencing it as a large object housing complex processes and producing powerful unusual sounds is meant to make the exhibit memorable to the visitors.

Visitors, usually in groups of two or three people, are free to explore the spaces in any order: start with an introduction and then proceed to the display or first interact with the exhibit and then ask questions and get more context or skip any of the parts. The audio-visual display supports simultaneous activation of any number of strips allowing a small group of visitors to interact with the exhibit together and experiment with combining sounds of different shapes. Screen recordings of two strips activated simultaneously are available online: ellipse (panned more to the left channel compared to the exhibit) + wave\footnote{\href{https://youtu.be/tTI19KiyDo0}{https://youtu.be/tTI19KiyDo0}}, ellipse (panned more to the left channel compared to the exhibit) + rectangular strip (panned more to the right channel compared to the exhibit) \footnote{\href{https://youtu.be/JPfCqN6OlNY}{https://youtu.be/JPfCqN6OlNY}}. In considerations of hygiene and visitor safety foot pedals are selected as a means of interaction with the display. Additionally, the microscope is equipped with a digital camera transmitting the image to a monitor allowing visitors to see the size comparison between the microstrip and a human hair without the need to touch the microscope. 

\begin{figure}
    \centering
	\includegraphics[width=0.45\textwidth]{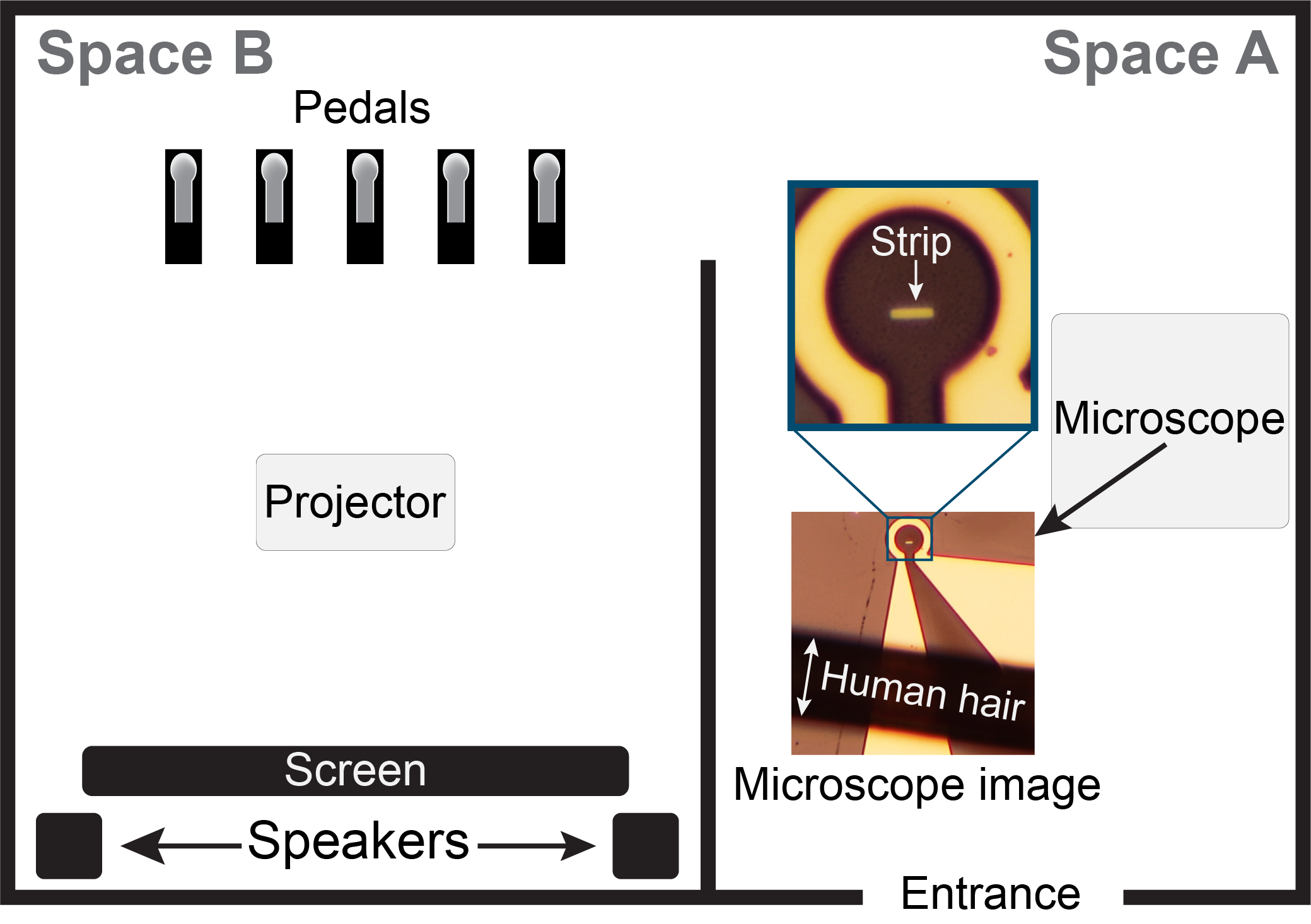}
	\caption{Plan of the Spin Wave Voices exhibit.}
	\label{fig:cubePlan}
\end{figure}

\section{Reception and Discussion}
\label{sec:page}
The installation, as described in Section~\ref{sec:Installation}, including both spaces and interactive guidance provided by the authors of this paper, was presented at an annual international digital art exhibition Ars Electronica Festival in 2022 with the topic of the year \enquote{Welcome to Planet B}. There \textit{Spin-Wave Voices} introduced visitors to spin waves as a potential way to make computational devices more energy efficient and, thus, contribute to solving the global energy crisis. Additionally, a simplified version of the developed prototype was demonstrated at two research groups, one of which specializes in micromagnetic simulations, and another in research, which also includes measurements of complex magnetization configurations within magnetic materials, and at the INTERMAG 2024 as a contributed talk. The goal was to gather feedback from the specialists in the field.

\subsection{Ars Electronica Festival}
The festival spanned over 5 days and attracted 71,000 visitors at 11 locations situated in Linz, Austria\footnote{\href{https://ars.electronica.art/mediaservice/en/2022/09/12/71000-visits-to-planet-b/}{https://ars.electronica.art/mediaservice/en/2022/09/12/71000-visits-to-planet-b/}}.
\textit{Spin-Wave Voices} was placed in a glass cubicle situated in a hall with other exhibits at the Johannes Kepler University Linz campus.
In order to create dark space inside and provide sound insulation the cubicle was covered with thick dark stage curtains around its perimeter from the inside. The same kind of curtain was used to separate the Space~A with the microscope and the Space~B with the audio-visual display the way that the Space~B was not visible from the installation entrance (see Fig.~\ref{fig:cubePlan}). One of the reasons for the separation curtains placement was to prevent additional light in the Space~B, another reason was to create a maze-like feeling of the exhibit. The entrance to the cubicle was always open in order to invite visitors in and provide additional light to the room with the microscope\footnote{\href{https://youtu.be/DVGgFnQUK1g}{https://youtu.be/DVGgFnQUK1g}}. 

While giving an interactive introduction to the exhibit to all willing participants, the authors did their best to observe their reactions to the installation and collect verbal feedback. As no means of formal evaluation was available at the moment of the exhibition, below we list observations all five authors found relevant.

\begin{itemize}
    \item Placement of the separation curtain concealed the appearance of the Space~B from passing by visitors, only the part with a microscope, poster, and cleanroom suits was visible. This motivated more curious visitors to go inside and explore. In most cases, the curiosity was triggered by the AD, which was well audible from the outside and often attractive with its unusual sound. With the AD inactive, sometimes the exhibit was passed unnoticed.
    \item Rare visitors were repulsed by the sound, probably finding it unpleasant or too loud. However, of the people who interacted with the exhibit, none expressed any dissatisfaction with the volume level.
    \item Many visitors were willing to first find out more about the exhibit and its meaning before interacting with the audio-visual display. Apparently, the variety of objects present in the Space~A, the microscope, the cleanroom suits, and the scientific poster with more detail on the ongoing research of spin waves, helped engage different kinds of participants. The introduction part ended up being key to the exhibit both to convey the main messages of the exhibit but also to engage the visitors with the narrative.
    \item Majority of the visitors quickly grasped the concept of controlling the display with foot pedals. Attendees often took time to thoroughly explore the sound of each microstrip and their combinations. Some kept the shapes active far beyond the transient behavior enjoying the meditative experience provided by the pulsating visualization combined with the nuanced AD.
    \item Some visitors enthusiastically shared their appreciation of the visualization. Nevertheless, some visitors got confused by it. Visitors sometimes interpreted the ridge lines of the plots as individual waves. In particular, some visitors thought that the spin waves were waves that expand over a 1D line and were unable to comprehend the complex process of spin waves expanding over the entire plot. Hence, using ridge lines to avoid occlusion in the 3D plots introduced unwanted ambiguities.
    \item Noticeable amount of people got involved into a discussion about computational devices and, in particular, spin waves, and how the latter can make computational technologies more energy-efficient. People got interested in more details in what spin waves are and often asked a permission to make a photo of the poster to research further. That included not only general public, but as well physicists from other research fields. In some cases, visitors, who have chosen to have an introduction first and then an interaction with the audio-visual display (see Section~\ref{sec:Installation}), came back after the interaction to find out more about the phenomenon.
\end{itemize}

From the observations listed above the following conclusion can be drawn: the first three objectives listed in Section~\ref{sec:objectives} were reached, namely the public awareness of the need for energy-efficient computational technologies, e.g. spin waves was raised  (see objective (1) in Section~\ref{sec:objectives}); the visitors got an opportunity to dive deeper into nano-scale ultrafast processes akin to ones behind modern and future computational technologies (see objective (2) in Section~\ref{sec:objectives}); and an accessible immersive audio-visual experience was provided, which illustrates the basic concept of spin waves in computing (see objective (3) in Section~\ref{sec:objectives}). The fourth objective to explore the potential of sonification as a tool for analysis of spin-wave data is discussed in Section~\ref{sec:demoSpecField}). 

Given that the conclusions stem from informal discussions with the visitors, in the future work it is planned to conduct formal evaluation of the setup to quantify participant initial interest, engagement and learning effect (e.g. through monitoring of objective parameters, such as time spent interacting with each part of the exhibit, as well as via questionnaires). Nevertheless, the listed observations have already shown that the following improvements might help to get an even better reach in the future: (1) an audio-based exhibit needs ways to attract the public when the setup is not in action, (2) artistic and science communication objectives can clash, for example when visitors do not engage with the message of the exhibit because of its looks or sound being alien to them, (3) providing explicit interactive introduction can increase visitor engagement with the exhibit, (4) special care needs to be taken to prevent potential misinterpretation of a science communication exhibit.

\subsection{Demonstration to Specialists in the Field}
\label{sec:demoSpecField}
The idea of \textit{Spin-Wave Voices} with short demonstration was presented in two research groups, one of which specializes in micromagnetic simulations, and another in research, which also includes measurements of complex magnetization configurations within magnetic materials. Additionally, similar presentation was done at the INTERMAG 2024 as a contributed talk. The idea of the project was suggested as a potential augmentation of visual analysis of complex spin-wave data. The feedback from the specialists is summarized in the following list:
\begin{itemize}
    \item The prototype allowed us to investigate different spin-wave patterns simulated in the systems with different shapes of the strips. With the help of the sonification, it was possible to recognize without any additional Fourier analysis multiple spin-wave eigenmodes in all shapes, except for the ellipse [see Fig.~\ref{fig:shapes}~(b)], where at the chosen excitation parameters the uniform excitation is observed.
    \item Some of the specialists were excited to try even the current implementation on their data to maybe get some new insights about it. But with the current implementation (see Section~\ref{sec:sonAD}) that was not possible during the display.
    \item Researchers dealing with complex 3D vector fields were enthusiastic to develop a sonification, which would allow them to search for certain phenomena in their data, as visualization tools are limited in that regard.
    \end{itemize}
Therefore, the objective (4) in Section~\ref{sec:objectives}, namely to explore the potential of sonification as a tool for analysis of spin-wave data, was reached as well. If developed further, the approach can find its application in the magnonics research field and further, in the research of the complex magnetic phenomena. With the little improvement, e.g. allowing to easily switch the sonified data and adjust the parameters of the visualization and, hence, AD, even the current approach could already be used as a tool for spin-wave data analysis.

\section{Conclusion and Outlook}
This paper introduces an approach to constructing an audio-visual representation of complex spin-wave data. The work at hand showcases the use of the approach as a part of a science communication art exhibit and outlines steps to be taken to develop it into an efficient tool for the analysis of spin-wave behavior, allowing researchers to rely both on their vision and hearing for discovering new patterns in the data. \textit{Spin-Wave Voices} featuring the presented approach was successfully demonstrated at the annual international digital art exhibition Ars Electronica Festival 2022 \enquote{Welcome to Planet B}, where it has shown potential to communicate recent advances in spin-wave research, which aims for more energy-efficient computational devices, to general public (see objective (1) in Section~\ref{sec:objectives}), served as an engaging illustration to spin waves as a process beyond human perception (due to its scale and speed) (see objective (2) in Section~\ref{sec:objectives}) and provided a captivating audio-visual experience to the visitors (see objective (3) in Section~\ref{sec:objectives}). The paper reports details of the setup as well as observations on the visitor reception of the exhibit. Additionally, results of the presentation to the specialists in the magnetic phenomena research are discussed. This served as an initial proof of concept encouraging further investigation of ways to leverage sonification for spin-wave data analysis (see objective (4) in Section~\ref{sec:objectives}).

Given that the conclusions presented in this paper stem from informal discussions with the visitors of the digital art festival and the physicists, in the future work it is planned to conduct formal evaluation of the setup as a science-communication tool with the main objective to quantify its attractiveness and effectiveness. The results of the evaluation will be the guiding factor to improving the exhibit: making sure it conveys intended messages, mitigating interaction bottlenecks and boosting accessibility. The next steps in the development of the exhibit based on the presented audio-visual display include making it more engaging and informative and expanding its scope from general public to schools, the media, policy makers, as well as to the scientific community. Boosting visitors engagement could be achieved by introducing an attractive stand-by mode and giving visitors more control over the installation, e.g. choosing base frequencies and other sonification parameters for each strip. In turn, addressing visualization ambiguity and demonstrating spin waves in comparison to modern technology, e.g. showcasing differences in produced heat through engaging thermoreception of the visitors, could make the exhibit more informative. Another direction of future work is to develop the approach into an independent tool for spin-wave data analysis. This would require unifying the code/software base, e.g. turning the display into a single Python application that would allow for easier deployment and integration with other tools. A study would be required to determine the most useful modes of sonification, each aimed at its own analysis task. Finally, the tool will need to provide flexibility in terms of sonification: manual path selection, cropping of the data, and area aggregation parameters.

\section{ACKNOWLEDGMENT}
\label{sec:ack}
The authors thank A.~Ney, V.~Ney, and M. Streit for their support during the early stage of the project. Also, the authors would like to acknowledge funding by Johannes Kepler University Linz, Linz Institute of Technology (LIT), the State of Upper Austria, the Federal Ministry of Education, Science, and Research (LIT-2019-7-SEE-117 and LIT-ARS-2022-009), the FWF Austrian Science Fund (DFH 23–N and ESP 4), and the European Research Council (ERC) under the European Union’s Horizon 2020 research and innovation program (101019375).

\printbibliography

\end{sloppy}
\end{document}